\documentclass[conference]{IEEEtran}
\pdfoutput=1
\makeatletter
\def\ps@headings{%
\def\@oddhead{\mbox{}\scriptsize\rightmark \hfil \thepage}%
\def\@evenhead{\scriptsize\thepage \hfil \leftmark\mbox{}}%
\def\@oddfoot{}%
\def\@evenfoot{}}
\makeatother
\usepackage{amssymb}
\usepackage{graphicx}
\usepackage{tikz}
\usepackage{cite}
\usepackage{amsmath}
\usepackage{mathtools}
\usepackage{hyperref}
\usepackage{epstopdf}
\usepackage[utf8]{inputenc}
\usepackage{pgfplots,pgfplotstable}
\begin{document}

\title{An Empirical Model of Packet Processing Delay of the Open vSwitch}
\author{\IEEEauthorblockN{\dag Danish Sattar and \ddag Ashraf Matrawy\\}
\IEEEauthorblockA{
\dag Department of Systems and Computer Engineering,
\ddag School of Information Technology\\
Carleton University,
Ottawa, ON Canada\\
Email: \{danish.sattar, ashraf.matrawy\}@carleton.ca}}
%\IEEEspecialpapernotice{\textcolor{red}{Draft for review version 8.0 - \today}}
\maketitle

\begin{abstract}
Network virtualization offers flexibility by decoupling virtual network from the underlying physical network. Software-Defined Network (SDN) could utilize the virtual network. For example, in Software-Defined Networks, the entire network can be run on commodity hardware and operating systems that use virtual elements. However, this could present new challenges of data plane performance. 
In this paper, we present an empirical model of the packet processing delay of a widely used OpenFlow virtual switch, the Open vSwitch. In the empirical model, we analyze the effect of varying Random Access Memory (RAM) and network parameters on the performance of the Open vSwitch. Our empirical model captures the non-network processing delays, which could be used in enhancing the network modeling and simulation. 
\end{abstract}
\IEEEpeerreviewmaketitle

\section{Introduction}

SDN is a new concept of networking with emphasis on better and easier network management. In traditional networks, each device is fully or partially autonomous in decision making and forwarding of information (packets)~\cite{Xiong2016172}. On the contrary, in SDN the control plane is solely responsible for making the forwarding decisions and the data plane only forwards the packets according to the rules set by the control plane. The control and data plane communicate using OpenFlow protocol.

SDN utilizes several virtual network components (e.g. switches, firewalls, intrusion detection/prevention systems, etc.) and one of the widely used virtual network switch used by the SDN is the Open vSwitch (OVS).  
In SDN, the network manager defines the traffic forwarding rules at the controller. Every new packet follows the following procedure to reach the destination node: Once a packet reaches the \texttt{in-port} of OVS, it checks whether it has a flow rule or not. If it has a flow rule, then forward the packet accordingly otherwise pass it to the controller. Next, the controller will check if there is a pre-defined rule for the incoming flow, if there is a rule then forward it using OpenFlow protocol to the OVS otherwise use broadcast to find a route in the network.

There are three types of delays experienced by a packet in the process above; i) controller delay: the time it takes to find an appropriate rule/route, ii) two-way propagation delay between the controller and the requesting OVS and iii) processing delay of the OVS.

We are motivated by the lack of studies, models, and simulators that consider the processing delays that happen in inside network elements. 
In this paper, we are investigating different processing delays experienced by the packet \textit{inside} the OVS to develop an empirical model for processing delays inside the OVS. The controller and two-way propagation delays are outside the scope of this paper. We have developed an empirical model to characterize these delays. We use commodity hardware to get the measurement in our empirical model. We performed experiments on two platforms; i) virtual and ii) baremetal. We tested using a variable amount of RAM and different network parameters on various hardware configurations. We believe that our empirical model can be beneficial for data plane simulators, and it could provide insight into choosing the appropriate platform (virtual or baremetal) for an application. 

We note that while we use different systems in our experimental setup, we are not trying to benchmark or test the performance of these systems or create comparisons with other systems. Rather, we are trying to capture the processing delay behavior of OVS in our experimental setup, and we understand that the behavior might be different in a different setup.

The rest of the paper is organized as follows.
In section~\ref{label:RelatedWork} existing works are presented.  Section~\ref{label:OVS} describes the Open vSwitch components and functionality. The empirical model is presented in section~\ref{label:EMOVS}. Results are discussed in section~\ref{label:VOI}, \ref{label:BOI} and finally section~\ref{label:conclusion} concludes this paper.

\section{Related Work}
\label{label:RelatedWork}
S. Azodolmolky \emph{et al.}~\cite{Azodolmolky2013} developed a network calculus based an analytical model to describe the functionality of Software-Defined Networks. They modeled the behavior of an SDN switch in terms of delay, queue length boundaries, buffer length and controller buffer length. K. Mahmood \emph{et al.}~\cite{Mahmood2015} provided SDN modeling based on queuing theory, where a Jackson network was used to model the data plane and \emph{M/M/1} queue was used to model the controller. They determined the average time a packet spends in the SDN and the maximum data that can be injected in the network given some delay requirements. Authors built a custom simulator to validate their analytical model. Another queuing theory based model was developed by Xiong\emph{et al.}~\cite{Xiong2016} to evaluate the OpenFlow-based software-defined network. To obtain the average time a packet spends in the system, an $M^X/M/1$ queue was used to model the switch. They evaluated the switch queuing model with different performance parameters using numerical analysis. The \emph{M/G/1} queue was used to model the controller \texttt{packet-in} behavior and it was evaluated using widely used benchmark Cbench under various network scenarios.

U. Javed \emph{et al.} developed a stochastic model for transit latency in SDN~\cite{Javed2017218}. They performed experiments on three different platforms (i.e. Mininet, MikroTik RouterBoard 750GL and GENI) and used the Round Trip Time (RTT) between end hosts as a measurement metric to formulate their model. They also proposed and demonstrated that the log-normal mixture distribution is more suited for transit latency in SDN as compared to M/M/1 models suggested in earlier studies. 
A hybrid approach was proposed by M. Jarschel \emph{et al.}\cite{Jarschel2011}, where they first used hardware switches to measure the average packet forwarding time then selected one of the hardware switch performance values to develop a queuing model to analyze the network. They simplified the OpenFlow architecture queuing system to \emph{M/M/1} (forwarding model), \emph{M/M/1-S} (controller model) from \emph{M/GI/1} and \emph{M/GI/1-S}, respectively. To validate the results from analytical model, they used OMNeT++ to implement packet based simulation.

\begin{figure*}[!ht]
    \centering
    \includegraphics[width=13cm,keepaspectratio=true]{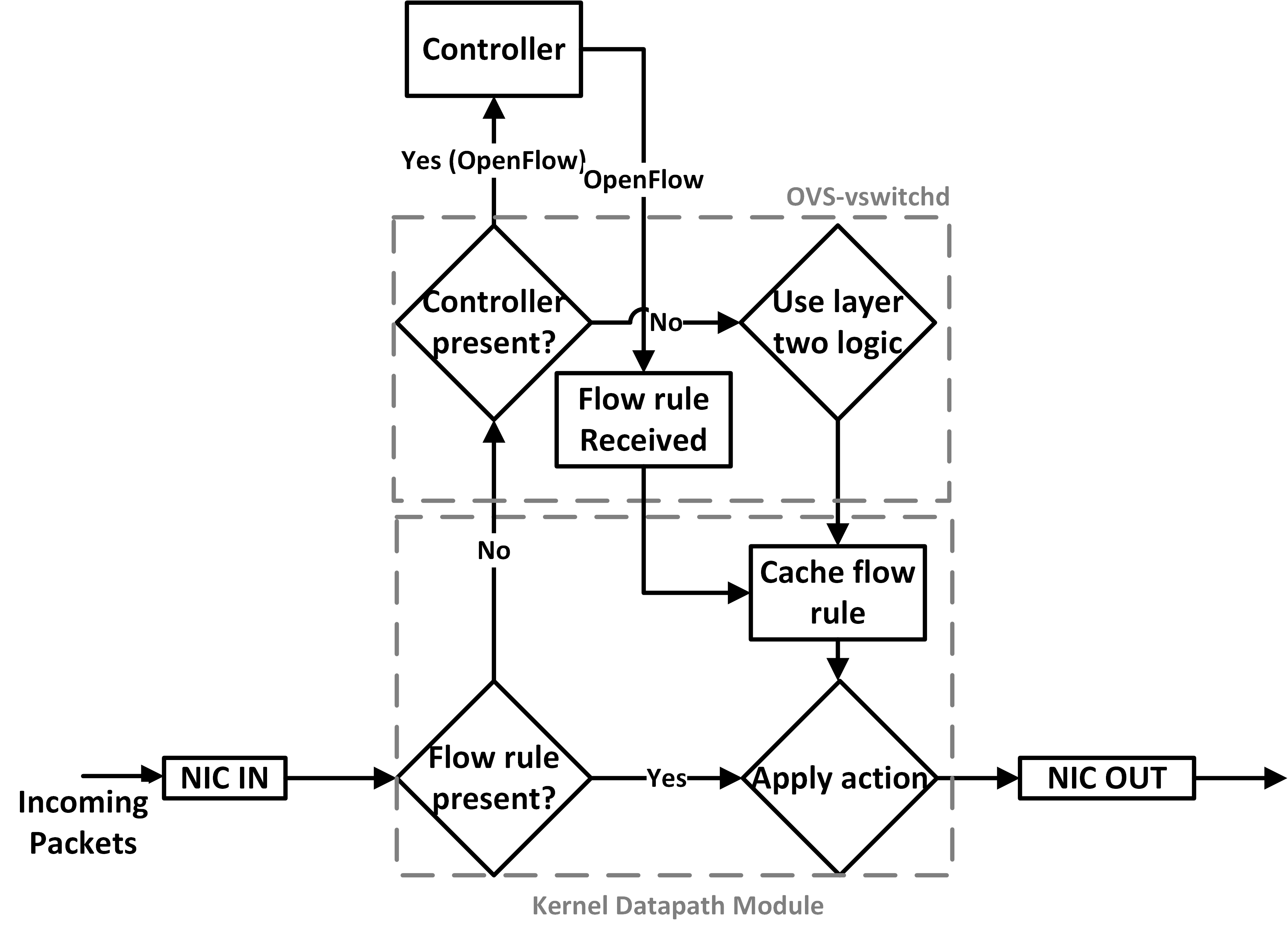}
    \caption{Overview of the working of the Open vSwitch}\label{fig:ovsint}
\end{figure*}

\section{Open vSwitch}
\label{label:OVS}
Open vSwitch (OVS) is a widely used OpenFlow virtual switch that is flexible and programmable. The OVS can work with or without a controller. In the first case, when there is a controller it uses the control logic to forward the packets. In the latter, if there is no controller present, it forwards the packet according to the layer two logic~\cite{ovsdoc}.

There are two major components of the OVS. The first is a userspace module called \texttt{ovs-vswitchd}. It is a userspace daemon that is implemented independently of the operating system. The second major component is \emph{kernel datapath module}, which varies operating system to operating system~\cite{188960}.

Figure~\ref{fig:ovsint} shows an overview of the working of the OVS. The \emph{datapath module} in the kernel always receives the packets from the Network Interface Card (NIC). Either \emph{kernel datapath module} or \texttt{ovs-vswitchd} has the instructions on how to handle this type of packet. In the first case, instructions called \emph{actions} are already cached at the \emph{kernel datapath module}, and it will forward the packets accordingly. In the latter, when the \emph{kernel datapath module}  does not have the actions cached, it will pass the packet to the \texttt{ovs-vswitchd}. Once a packet reaches the \texttt{ovs-vswitchd}, if there is a controller present and \texttt{ovs-vswitchd} does not have the instructions on how to handle these type of packets, it will forward it to the controller and wait for a reply. In the second case, when there is no controller present, it will use its internal logic to define actions for the incoming packets. Once the \texttt{ovs-vswitchd} has defined the appropriate actions, it will pass them to the \emph{kernel datapath module} and also instruct it to cache them for future use~\cite{188960}.

As mentioned before, the most common use of the OVS is as an SDN switch to control packet forwarding in OpenFlow. The OpenFlow protocol allows a controller to dynamically add, update, remove, obtain statistics on the flow tables and monitor the flows as well as inject, redirect or drop the packets. The \texttt{ovs-vswitchd} receives the flow tables from the SDN controller, matches any incoming packets against these OpenFlow tables, gathers the actions applied, and caches the results in the \emph{kernel datapath module}. It allows the OVS to work independently of any SDN controller as it only needs to understand the OpenFlow protocol. The separation of userspace module and kernel module is transparent to the OpenFlow protocol, which makes it simpler from the network programmer and SDN controller's point-of-view. In SDN controller's point-of-view, every packet goes through a series of OpenFlow tables and the OVS finds the highest priority matching flow whose conditions are satisfied by the packet and executes its OpenFlow actions~\cite{188960}.

\section{Empirical Model of OVS Processing Delay}
\label{label:EMOVS}
In the literature, there are several analytical models based on network calculus and queuing theory~\cite{Xiong2016172,Azodolmolky2013,Mahmood2015,Jarschel2011,Xiong2016,Bianco2010} to evaluate the performance of SDN. Most analytical models simplify the network elements to provide a mathematical formulation to evaluate the network performance but in reality, today's network is affected by several external factors as well. In recent years, a significant amount of network functionality has been virtualized (NFV), e.g., virtual routers, switches, firewalls, etc. These virtualized network functions run on top of an operating system (OS) or a hypervisor, which also affects the network performance.

In this paper, we provide an empirical model of the Open vSwitch to characterize different processing delays (non-network related delays) which affect the performance of the OVS. We used different sizes of RAM as well as variable packet sizes and data rates to cover a broad spectrum of configurations.
% We repeated the experiments on various hardware configurations several times to increase the accuracy of our measurements. 

In the packet processing function of \emph{kernel datapath module}~(Figure \ref{fig:ovsint}), each packet passes through four steps, which are:
\begin{itemize}
    \item get statistics of the current datapath from the CPU
    \item flow rule lookup in the flow tables
    \item if flow rule not found in the existing flow tables send it to the  \texttt{ovs-vswitchd} (upcall)
    \item update the statistics and apply actions
\end{itemize}
First, the datapath module gets some statistics (counters) about this datapath from the CPU then checks whether it has a cached flow rule or not. If it is present, it will use that rule otherwise send it to the userspace module (upcall) and lastly, it updates the statistics about the current flow (usedtime, packet, bytes, tcp\_flag, etc.) and executes the corresponding flow actions. We present delays in microseconds at each of these steps as an input for our empirical model.

Our methodology to calculate the empirical model is as follows~\cite{dess}; initially, we collect processing time data from each experiment then calculate their frequencies. In the second step, we convert those frequencies into relative frequencies according to equation~\ref{eq:freq} and calculate commutative relative frequencies to produce the empirical Empirical Cumulative Distribution Function (ECDF). 
\begin{equation}
\label{eq:freq}
\mathtt{
f=\frac{PT}{N}}
\end{equation}
\texttt{PT} is the processing time of a packet, \texttt{N} is the total number of packet samples for an experiment and \texttt{f} is the relative frequency. 

To cover a broad spectrum of configurations, we used two environments for the experimentation. In the first, we installed OVS in a virtual environment. For the second, OVS was installed directly on the baremetal. The OVS installation for all configurations is single core. We created multiple scenarios. In each scenario, we fixed all the parameters except one and analyzed the effect of that parameter on the packet processing time of OVS, and each experiment is repeated several times. In the following two sections, we discuss the results of virtual and baremetal OVS installation.

\section{Virtual OVS Installation (VOI)}
\label{label:VOI}
In the virtual OVS installation, we used Xen server 7.1~\cite{xen} to create a virtual machine (VM) and installed Ubuntu server 14.0.4 LTS with OVS 2.5.1~\cite{ovs251}. The hardware specifications of the host system are 8.0GB RAM and Intel E5420@2.50GHz (4 cores). 

In this configuration, we analyzed the effect of RAM, packet size and data rate on the processing time of the OVS.

\begin{figure}[!ht]
    \centering
    \includegraphics[width=\linewidth,keepaspectratio=true]{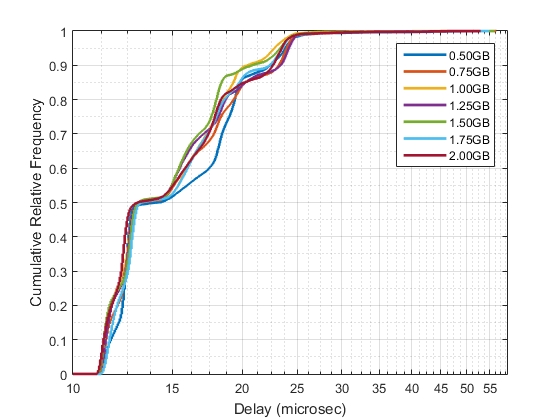}
    \caption{VOI: Variable RAM (ECDF).  \protect\linebreak RAM: 0.5GB-2.0GB, CPU: 1 core, Packet size: 56B and Data rate: 10-15Kb/s}\label{fig:xenRAM}
\end{figure}
\subsection{First scenario (variable RAM)} 
We analyzed the effect of RAM on the packet processing time by setting the number of CPU cores to 1, data rate to 10-15Kb/s and the packet size to 56B. Figure~\ref{fig:xenRAM} shows the ECDF of processing delay for different sizes of RAM (0.5GB to 2.0GB). Increasing the RAM does reduce the processing delay but it is not significant because the majority of packets experience the average total delay of approximately 25 $\mu s$ or less for all the RAM configurations.
\begin{figure}[!ht]
    \centering
    \includegraphics[width=\linewidth,keepaspectratio=true]{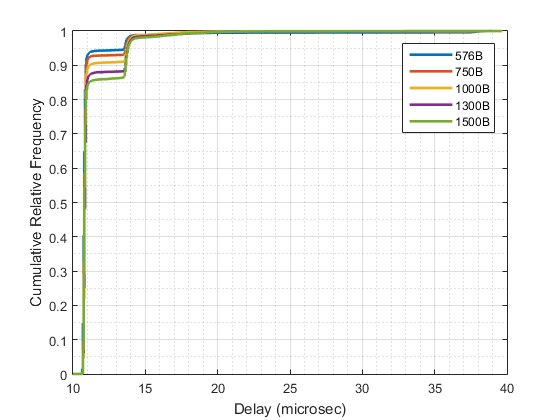}
    \caption{VOI: Variable data rate and constant packet size (ECDF).   \protect\linebreak RAM: 1.0GB, CPU: 1 core, Packet size: 576B and Data rate: variable}\label{fig:xenVDR}
\end{figure}

\subsection{Second scenario (variable data rate)} 
After varying the RAM in the first scenario, the RAM is set to constant values at 1.0GB for the second and third scenarios. 

We increase the packet size to 576B in the second scenario to allow for higher data rates. We present the effect of different data rates in Figure~\ref{fig:xenVDR}. At lower data rate (250 Kb/s), the OVS takes more time to process each packet. It is also evident in comparison with the previous scenario (Figure \ref{fig:xenRAM}), where data rates (10-15Kb/s) were much lower that Open vSwitch took more time to process a packet as compared to the higher data rates (compare the x-axis values of the Figures~\ref{fig:xenRAM} and \ref{fig:xenVDR}). U. Javed \emph{et al.}~\cite{Javed2017218} reported similar behavior while they considered the Round Trip Time (RTT) as a metric. They also reported how the higher data rates lead to lower processing delays. We believe this could be an explanation for the results we get for the processing delay with varying data rates, i.e., they are impacted by context switching and cache hits.

\subsection{Third scenario (variable packet size)} 
\begin{figure}[!ht]
    \centering
    \includegraphics[width=\linewidth,keepaspectratio=true]{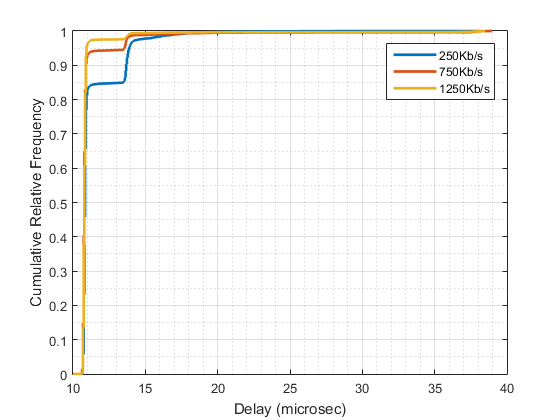}
    \caption{VOI: Variable packet size and constant data rate (ECDF).  \protect\linebreak RAM: 1.0GB, CPU: 1 core, Packet size: variable and Data rate: 750Kb/s}\label{fig:xenVPS}
\end{figure}

We fixed the data rate to 750Kb/s and used variable packet size. The smaller packets take less time to process but the processing time for larger packets is also not significantly different from smaller packet sizes as shown in Figure~\ref{fig:xenVPS}. The processing delay in this scenario is very similar to the second scenario (Figure~\ref{fig:xenVDR}), a maximum delay experienced by any packet is less than or equal to $ 40\mu s$. 

\subsection{Comparing different types of delay}

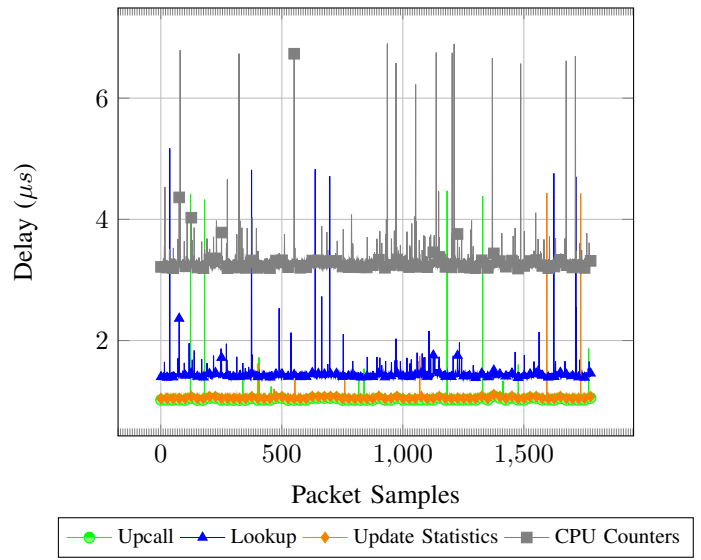
\begin{figure}[!ht]
    \centering
    % Preamble:
    %\pgfplotsset{width=\linewidth,compat=1.13}
    \begin{tikzpicture}[baseline]
    \begin{axis}[
    %ymode = log,
    %title=Inv. cum. normal,
    xlabel={Packet Samples},
    ylabel={Delay ($\mu s$)},
    minor x tick num=50,
    grid=major,
    legend style={font=\footnotesize,at={(0.5,-0.19)},
        anchor=north,legend columns=-1},
    ]
    \addplot[green,mark color=green!50!white,mark=halfcircle*,mark repeat=25,mark size=2pt] table {xenifelse.txt};
    \addplot[blue,mark color=blue!50!white,mark=triangle*,mark repeat=25,mark size=2pt] table {xenlookup.txt};
    \addplot[orange,mark color=orange!50!white,mark=diamond*,mark repeat=25,mark size=2pt] table {xenstats.txt};
    \addplot[gray,mark color=gray!50!white,mark=square*,mark repeat=25,mark size=2pt] table {xencounter.txt};
    \legend{Upcall,Lookup,Update Statistics,CPU Counters}
    \end{axis}
    \end{tikzpicture}
    \caption{VOI: Processing delay comparison of different components of \emph{kernel datapath module}} \label{fig:xenresponse}
\end{figure}

One interesting result we obtained from our experiments is the comparison of various components of packet processing delay inside the \emph{kernel datapath module}. In particular, we refer to the four delays(i.e. upcall, lookup, statistics update and CPU counters). Figure~\ref{fig:xenresponse} captures the delays of each component. The parameters used for this experiment are same as the third scenario. The OVS spends a considerable amount of time (in comparison with other types of delay) obtaining information (or waiting for the CPU) about the current datapath from the CPU.  The second largest component that contributes to the overall processing delay of the OVS is flow lookup delay.

We also noticed that the time required to obtain CPU counters is higher then upcall to the userspace daemon, flow lookup and updating the datapath statistics (and preform necessary flow actions). 

We also performed experiments where we added 2000 arbitrary flow rules to OVS's flow tables to see the effect on the lookup delays (not shown here). The flow lookup delays we obtained in these experiments were not as significant as we expected.

\section{Baremetal OVS Installation (BOI)}
\label{label:BOI}
In the second configuration, we installed Ubuntu server 14.0.4 LTS with OVS 2.5.1 directly on the baremetal. The second configuration is used to provide results for the different network parameters. The hardware specifications of the baremetal system are 8.0GB RAM and Intel Core 2 Quad Q6600@2.4GHz (4 cores).  
In this configuration, we variated the packet size and data rate. The network parameters for both scenarios are similar to VOI experiments (section \ref{label:VOI}). We also performed these experiments on multiple baremetal systems that produced results (not shown here) similar to BOI results.  

\begin{figure}[!ht]
    \centering
    \includegraphics[width=\linewidth,keepaspectratio=true]{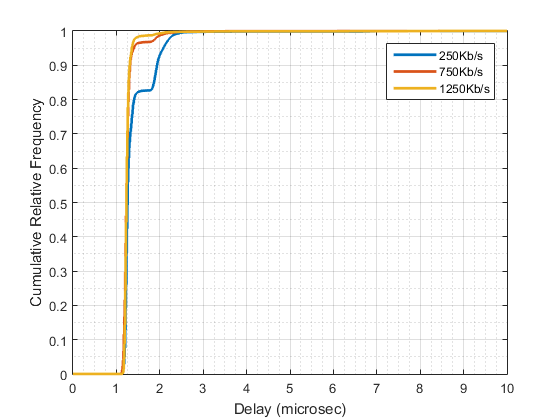}
    \caption{BOI: Variable data rate and constant packet size (ECDF).   \protect\linebreak RAM: 8.0GB, CPU: 4 cores, Packet size: 576B and Data rate: variable}\label{fig:bareVDR}
\end{figure}

\subsection{First scenario (variable data rate)} 
We repeated the variable data rate experiment on baremetal as shown in Figure~\ref{fig:bareVDR}. On baremetal, OVS produced similar behavior to the VOI, but it did improve the processing time of the OVS. The processing time is reduced to $10 \mu s$ from $40 \mu s$ (Figure~\ref{fig:xenVDR}).

\subsection{Second scenario (variable packet size)}
\label{label:second}
\begin{figure}[!ht]
    \centering
    \includegraphics[width=\linewidth,keepaspectratio=true]{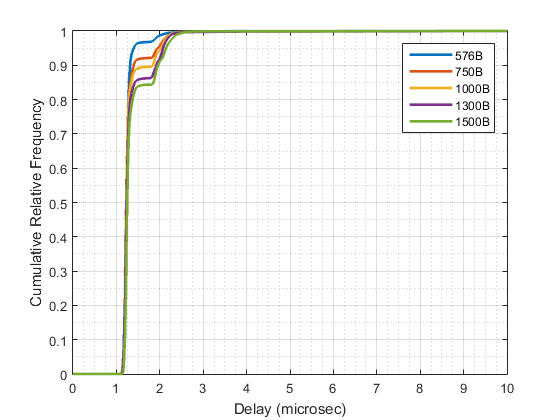}
    \caption{BOI: Variable packet size and constant data rate (ECDF).  \protect\linebreak RAM: 8.0GB, CPU: 4 cores, Packet size: variable and Data rate: 750Kb/s}\label{fig:bareVPS}
\end{figure}

In the second scenario, we repeated the variable packet size experiment on baremetal. This experiment also produced similar results as shown in Figure~\ref{fig:bareVPS}. The OVS took more time to process larger packets but an interesting point to note here is that in VOI the minimum processing delay experienced by any packet is approximately $10 \mu s$. On the other hand, $10 \mu s$ is the maximum delay encountered by any packet in the BOI, which is a significant reduction as we expected. 

\subsection{Comparing different types of delay}
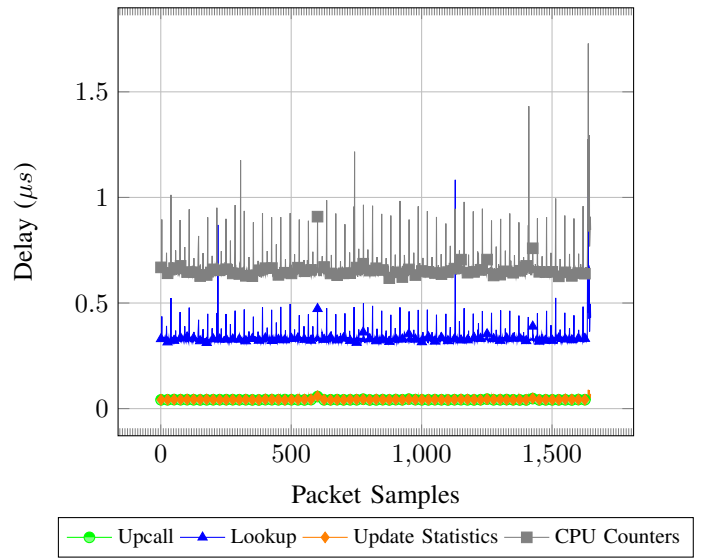
\begin{figure}[!ht]
    \centering
    % Preamble:
    %\pgfplotsset{width=\linewidth,compat=1.13}
    \begin{tikzpicture}[baseline]
    \begin{axis}[
    %ymode = log,
    %title=Inv. cum. normal,
    xlabel={Packet Samples},
    ylabel={Delay ($\mu s$)},
    minor x tick num=50,
    grid=major,
    legend style={font=\footnotesize,at={(0.5,-0.19)},
        anchor=north,legend columns=-1},
    ]
    \addplot[green,mark color=green!50!white,mark=halfcircle*,mark repeat=25,mark size=2pt] table {ifelse.txt};
    \addplot[blue,mark color=blue!50!white,mark=triangle*,mark repeat=25,mark size=2pt] table {lookup.txt};
    \addplot[orange,mark color=orange!50!white,mark=diamond*,mark repeat=25,mark size=2pt] table {Stats.txt};
    \addplot[gray,mark color=gray!50!white,mark=square*,mark repeat=25,mark size=2pt] table {wait.txt};
    \legend{Upcall,Lookup,Update Statistics,CPU Counters}
    \end{axis}
    \end{tikzpicture}
    \caption{BOI: Processing delay comparison of different components of \emph{kernel datapath module}} \label{fig:bareresponse}
\end{figure}

We used same network parameters as scenario~\ref{label:second}. This experiment also produced similar behavior. The CPU counters contributed the most in the OVS processing time followed by the lookup delay as shown in Figure~\ref{fig:bareresponse}. One difference we can see between Figure~\ref{fig:xenresponse} and \ref{fig:bareresponse} is that in BOI all delays are more consistent and stable as compared to VOI. The reason could be that in BOI, the CPU scheduling is done by the OS. While in VOI there are two levels of scheduling; one by the OS and other by the Xen Server.

\section{Conclusion}
\label{label:conclusion}
An empirical model of Open vSwitch has been presented in this paper. We conducted experiments using a different hardware and network configurations. Our empirical model captures the non-network processing delays that a packet experiences due to the nature of different OVS installations as well as other operating system factors. These experiments revealed some interesting results. Increasing the RAM size did not have a significant effect on the delay a packet experiences inside the OVS in the experiments we ran. Smaller packet size and higher data rates reduce the overall processing time of the OVS. 

\section*{Acknowledgement}
The second author acknowledges funding from Canada’s NSERC thought the Discovery Grant Program.

\bibliographystyle{IEEEtran}
\bibliography{EM_bib}
\end{document}